# Manipulating Interlayer Excitons for Ultra-pure Near-infrared Quantum Light Generation

Huan Zhao[1,*], Linghan Zhu[2], Xiangzhi Li[1], Vigneshwaran Chandrasekaran[1], Jon Kevin Baldwin[1], Michael T. Pettes[1], Andrei Piryatinski[1,3], Li Yang[2,4,*], Han Htoon[1,*]

[1]Center for Integrated Nanotechnologies, Materials Physics and Applications Division, Los Alamos National Laboratory, Los Alamos, New Mexico 87545, USA

[2]Department of Physics, Washington University in St. Louis, St. Louis, Missouri 63130, USA

[3]Theoretical Division, Los Alamos National Laboratory, Los Alamos, NM 87545

[4]Institute of Materials Science and Engineering, Washington University in St. Louis, St. Louis, Missouri 63130, USA

[*]Corresponding authors, Email: huanzha@lanl.gov, lyang@physics.wustl.edu, htoon@lanl.gov

## Abstract

Interlayer excitons (IXs) formed at the interface of atomically-thin semiconductors possess various novel properties. In a parallel development, nanoscale strain engineering has emerged as an effective means for creating 2D quantum light sources. Exploring the intersection of these two exciting areas, where strain and defects are exploited for the manipulation of IX toward the emergence of new functionalities, is currently at a nascent stage. Here, using MoS$_2$/WSe$_2$ heterostructure as a model system, we demonstrate how strain, defects, and layering can be utilized to create defect-bound IXs capable of bright, robust, and tunable quantum light emission in the technologically important near-infrared spectral range. We were able to achieve ultra-high single-photon purity with $g^{(2)}(0) = 0.01$. Our strategy of creating site-controlled QEs from the defect-bound IXs represents a paradigm shift in 2D quantum photonics research, from engineering intralayer exciton in monolayer structures towards IXs at the interface of 2D heterostructures.

## Main Text

### *Introduction*

Heterobilayers (HBLs) of transition metal dichalcogenides (TMDCs) with staggered type–II band alignment are particularly fascinating as they allow the formation of charge-separated interlayer excitons (IXs), where electrons and holes have their energy extrema in opposite layers (*1-5*). Recently, IXs have been emerging as an exciting ground not only for exploring fascinating many-body phenomena such as exciton condensation (*1-4*), but also for realizing exciton-based information processing technologies (*6, 7*). Formation of such IXs gives rise to photoluminescence (PL) emissions at the lower energy side. In most hetero-bilayers such as $MoSe_2/WSe_2$, momentum-direct IXs locates at K valleys of constituent layers (K-K IX, Fig. 1A) dominate the optical emission when the inter-layer twist angles are small (*5, 8-10*). Spatial charge separation in such IXs results in significantly reduced oscillator strength and an inefficient PL emission (*2*). On the other hand, in certain heterobilayers such as $MoS_2/WSe_2$ and $WS_2/MoS_2$, strong interlayer hybridization at the $\Gamma$ valley leads to $\Gamma$-K transition (Fig. 1B) with enhanced oscillator strength (*11-17*). However, bright PL emission is still hindered by its momentum-indirect nature.

While the emission properties of both momentum-direct K-K IX (emitting at ~ 1.0 eV) and momentum-indirect $\Gamma$-K IX (emitting at ~ 1.57 eV) have been investigated for $MoS_2/WSe_2$ HBLs (*16, 18, 19*), effects of strain field and in-gap defect states on complex IX transitions remain to be understood. Particularly, by band structure engineering, an electron residing in the sulfur-vacancy state ~200 meV below the MoS$_2$ conduction band edge (*20*) could be utilized to form optically active excitonic transition with a hole in the hybridized Γ band (Fig. 1C). This defect-bound Γ-defect IX retains large oscillator strength of the hybridized Γ IX species while bypassing stringent momentum conservation limitations in traditional K-K IXs which require matched lattice constants and perfect rotational alignment among the two constituent layers. Owing to the deep trapping potential, this defect-bound, Γ-band IX state (defect-Γ IX) could enable high purity quantum light generation in technologically important yet rarely reached near-infrared (NIR) spectral range and raise the operating temperature out of cryogenic regime.

## *Results*

### *IX emission from MoS$_2$-WSe$_2$ heterostructures*

To explore this potential, we fabricate nanopillars made of dielectric material, with the diameter of 150 nm and height of 200 nm, directly on a gold substrate. We then prepared three categories of heterostructures on top of the nanopillars/gold substrate: (i) MoS$_2$/WSe$_2$ HBLs (also called HBLs for short, illustrated in Fig. 1D); (ii) MoS$_2$/WSe$_2$ HBL on top of a 5 nm-hBN layer (HBL/hBN, illustrated in Fig. 1E); (iii) bilayer-MoS$_2$/ monolayer-WSe$_2$ (also called hetero-trilayer, or HTL, illustrated in Fig. 1F). Here nanopillars not only induce local strain within MoS$_2$/WSe$_2$ heterostructures but are also prone to create point defects via structural damage (*21-24*). The hBN layer in sample category (ii) isolates the HBL from gold-induced fluorescence quenching and serves as a protective layer that shields the HBL from structural damages or substrate-induced adsorbates - hence defect generation - during the stamping process onto nanopillars. Since the Γ valley features an out-of-plane orbital character, its orbital hybridization is sensitive to interlayer interactions (*11, 17, 25, 26*). For this reason, we stacked MoS$_2$/WSe$_2$ heterostructures with near-zero or near-60° twist angle with ±5° error (unless otherwise specified) to achieve small layer separation (*16*) and strong interlayer interactions (*27, 28*). We developed a modified mechanical exfoliation method to efficiently create and transfer large (~ 100 µ$m$) heterostructures (Supplementary Section 1). Through Raman mapping, an average strain level of 0.27% was extracted from the Raman difference between the homogeneous HBL and the HBL on nanopillars (Supplementary Section 2).

Under widefield illumination at 750 nm wavelength, low-temperature PL images of the heterostructures were acquired by an InGaAs detector (Fig. 1H & Supplementary Section 3), revealing that NIR emitters are created at point-of-contact regions in HBL/nanopillars. Note that we use a gold substrate to quench the PL emission from the homogeneous HBL region and thus confine the IX emission at nanopillar sites. When switching excitation wavelengths from 750 nm to 660 nm (Fig. 1I), the HBL/hBN region (Fig. 1E and white box in Fig. 1G) becomes remarkably brighter while the HBL/nanopillars region (Fig. 1D and brown box in Fig. 1G) significantly dims (explained in detail later). Next, we spectrally resolve the HBL/nanopillar emitters and find that the ~1.0 eV broad PL peak of K-K IX commonly reported in $MoS_2$/$WSe_2$ HBLs is replaced by narrow PL lines emerging mainly in the 1.35±0.08 eV range (Fig. 1J, 1K), corresponding to a redshift distributed between 140 and 300 meV from the Γ-K IX emission energy at 1.57 eV(*16*). Since this energy redshift is in agreement with that of the sulfur-vacancy-induced PL redshift observed in $MoS_2$ monolayers (*20, 29*), we attribute the transition to Γ-defect IX formed between an electron trapped in the sulfur defect levels and a hole in the hybridized Γ band of the HBL (the case in Fig. 1C). For HTL/ nanopillars (Fig. 1F), the emission energy is further redshifted to 1.26±0.1 eV range without significant reduction in PL intensity, advocating that we can utilize layer engineering as a tuning knob to control the Γ-defect IX emission range. (See Supplementary Section 4 for more statistics and PL spectra).

Interestingly in HBL/hBN region, where the hBN serves as a shielding layer minimizing defect generation and gold-induced PL quenching, a complete suppression of the Γ-defect IX emission and recovery of the delocalized K-K IX transition at 1.0 eV is observed. Due to nanopillar strain, this emission exhibits a redshift and spectral broadening compared to that observed within unstrained regions (Fig. 1J). This is consistent with recent studies reporting that narrow-linewidth localized emissions cannot appear solely due to strain effect (*21, 30*). We also examined an HBL/nanopillar sample with 22° twist angle and found that the emission is significantly dimmer and shows broader linewidth (Supplementary Section 5). These findings provide evidence that control of twist angle, layer structure, defect concentration, and excitation wavelength can be utilized along with the nanoscale strain engineering to achieve localization and activation of various IX transitions (see Supplementary Table 1 for a summary).

## *Demonstration of Ultra-high purity single photon emission*

To demonstrate that the Γ-defect IX exciton states can act as quantum emitters (QEs), we performed time-tagged, time-correlated photon counting, and the Hanbury–Brown–Twiss (HBT) experiments at 10 K. Fig. 2A shows the PL spectrum of a typical emitter, of which the spectrally filtered emission is guided into a two-channel single-photon detector to obtain time-resolved PL (TRPL) curve (Fig. 2B), time-dependent PL counts (Fig. 2C), and second-order photon correlation ($g^{(2)}$) trace (Fig. 2D). Measured TRPL shows lifetime of 18.45±0.01 ns. Interestingly, the first few nanoseconds of the TRPL curve features a slow uprising from initial non-zero value. This is distinctly different from the TRPL curves of the K-K IX (Supplementary Section 7) showing monotonous decay, implying a delay between optical absorption and photocarrier recombination

(explained in detail later). The stability of the emission is furthered confirmed by time-dependent PL counting. The $g^{(2)}$ trace showing almost complete photon antibunching with $g^{(2)}(0) = 0.01$ (our experimental sensitivity limit) provides unequivocal confirmation of the single-photon nature of our Γ-defect IX. We note that the purity of this single-photon emission is significantly higher than that recently reported (*7*) in moire-trapped excitons ($g^{(2)}(0)$ ~0.28, see Supplementary Section 8 for a discussion of moire IX and the IXs reported in our work). Such an ultra-high single photon purity now meets the strict requirement for implementation of quantum key distribution (QKD), logic gates, and memory technologies (*31*). We attribute this dramatic improvement of single-photon purity to the fact that our QE is in complete spectral isolation from any of the intra- and inter-layer excitonic transitions and the gold substrate effectively quenches the emission of the HBL except directly on top of the dielectric nanopillars. HBT experiment performed on a different emitter at 77 K reveals that the single-photon emission can sustain up to liquid nitrogen temperature (Fig. 2E). The temperature-dependent PL experiment shows that while the PL lines get broader at elevated temperatures, the reduction of the integrated PL intensity is essentially insignificant (Fig. 2F).

*The physical origin of the QE*

Next, to confirm the Γ-defect IX states as physical origin of our ultra-pure QEs, we performed first-principles calculations on the electronic structures of the $MoS_2$/$WSe_2$ heterostructures. Compared with intrinsic HBL (Fig. 3A), the band structure of strained HBL (Fig. 3B) shows that strain brings the Γ band closer in energy to the $WSe_2$ K valence band, facilitating the $WSe_2$ hole transfer. When a sulfur vacancy is introduced, defect bands emerge below the conduction band minimum (CBM), in qualitative agreement with the red-shift of Γ-defect IXs (Fig. 3C). The introduction of Se vacancy, on the other hand, results in defect levels above the CBM, leaving the low-energy optical transitions unchanged (Supplementary Section 9). While the Γ-band is highly hybridized and has a sizable amount of $MoS_2$ component, the VBM at K only has contribution from the $WSe_2$ layer. This fact, together with the confinement of electron in the S vacancy level, allows Γ-defect IX transition to dominate over K-defect IX transition. Once an hBN layer is inserted between the HBL and nanopillars, defect formation in $MoS_2$ layer is averted as evidenced by the clean PL spectrum of the hBN encapsulated $MoS_2$ (Fig. 3E), and thus the K-K IX emission in the HBL is recovered. The room temperature PL measurements further confirm that the Γ-K emission in homogenous HBL is replaced by the Γ-defect emission in HBL samples on nanopillars (Fig. 3F). Compared to HBL, the calculation for a HTL composed of bilayer $MoS_2$ and monolayer $WSe_2$ reveals a reduced band gap and an increased $MoS_2$ component at Γ band maximum (Fig. 3D), leading to further redshifted NIR defect Γ-defect IX emissions (Fig. 1K and Fig. S5b) and enhanced $MoS_2$-branch PLE features (Supplementary Section 10).

*PL excitation spectroscopy*

Finally, to understand strong influence of excitation wavelength on the competition between K-K and Γ-defect IX emissions, we conducted PL excitation (PLE) experiments on three types of samples: (i) HBL/hBN/nanopillar sample for probing K-K IX; (ii) HBL/nanopillar and (iii) HTL/nanopillar samples for probing Γ-defect IXs. For K-K IX emission from HBL/hBN on nanopillars, we observed enhanced PL intensity when the excitation is in resonance with delocalized 2D intralayer excitons (i.e., $WSe_2$ exciton at 1.75 eV and $MoS_2$ A & B excitons at 1.88 eV and 2.01 eV, respectively) (Fig. 4A), evidencing that the K-K IX is primarily formed by nonlocal interband absorption in constituent layers followed by interlayer charge transfer (Fig. 4C). We also observed an absorption peak at 1.66 eV, which corresponds to the visible spectral range

PL emission of HBL on nanopillars stemming from the interplay of strain-induced bandgap narrowing and strain-activated dark exciton emission in the WSe$_2$ layer (Fig. 4E). In stark contrast, the Γ-defect IX emitting at 1.34 eV from HBL/nanopillar shows a series of PLE peaks near the band edges of MoS$_2$ and WSe$_2$, predominantly in the 1.5 to 1.7 eV range (Fig. 4B and Fig. S11a). Considering their resemblance to the emissions from WSe$_2$ localized excitons, we attributed the PLE features in 1.5 to 1.7 eV range to localized transitions that originate from the interband absorption of defect states in WSe$_2$ layer amid local strain (Fig. 4D). Such absorption patterns dominated by localized optical transitions were also observed in HTL/nanopillar samples (Fig. S11b) but were not detected in HBL/hBN/nanopillar structures. Compared to HBLs, HTLs shows enhanced PLE peaks near MoS$_2$ band edges due to increased MoS$_2$ component at Γ band edge (Fig. 3D). Further discussions of the PLE features are provided in Supplementary Section 10.

## *Discussion*

For Γ-defect IX emissions, HBL are coupled directly to the Au substrate except for the regions immediately above the nanopillars. As a result, only the excitons created in the nanopillar regions can contribute to the Γ-defect IX emissions while all the delocalized 2D intralayer excitons created in the vicinity of nanopillars were essentially quenched by gold. At nanopillar sites, excitons resonantly pumped into the 1.5 – 1.7 eV energy states are trapped by the strain-and-defect induced potential wells before relaxing to the Γ-defect IX state, leading to the prominent absorption features. For K-K IX emissions from HBL/hBN/nanopillars, the hBN spacer layer placed between the HBL and nanopillar arrays on gold substrate minimizes defect formation and suppresses the quenching of the 2D intra-layer excitons. Therefore, 2D intra-layer excitons formed in homogenous HBL/hBN regions nigh the nanopillars could be funneled into the nearest nanopillar sites and contribute to the delocalized-exciton-dominated PLE features of the K-K IX emission in absence of defect states. Taken together, these findings indicate that resonant excitation into WSe$_2$ localized exciton states as well as quenching of 2D intra-layer excitons play a crucial role in defect-Γ IX quantum light emissions.

The fact that Γ-defect IXs are mainly populated via photoexcitation in WSe$_2$ absorption bands explains the slow uprising feature observed in the first few nano-second of the TRPL curve (Fig. 2B). When the repetition rate of the pump laser is increased, the uprising feature becomes more pronounced due to the acceleration of carrier injection (Supplementary Section 11). Such a slow rise suggests slow arrival of holes to the hybridized Γ band. While the electron transfer from the CBM of WSe$_2$ to the S-vacancy state is energetically downhill suggesting fast electron transfer, the hole transfer from the K point of WSe$_2$ to the hybridized Γ point occurs uphill and requires an activation energy at low temperature. To get a better insight into the exciton photoexcited dynamics, a nonlinear rate equation model for coupled K/-K exciton state in WSe$_2$ and Γ-defect IX is introduced in Supplementary Section 11. The model successfully reproduces the initial time uprising feature in Fig. 2B and yields slow exciton transfer time ~5 ns to the defect state. A good agreement was achieved (Fig. S13a) between our theory and experiment giving strong evidence towards validity of the proposed above (Fig. 4D) electronic state model. Additionally, the model explains the decay of the side peaks of the g$^{(2)}$ function, i.e., the long-time bunching feature, in Fig. 2D, which becomes more pronounced when extending the delay time to µ$s$ timescale (Fig. S13b). We attribute this bunching behavior to a dark defect shelving state receiving some portion of population from the Γ-defect IX on the timescale of 100 ns and re-populating back on the timescale of 2 µs. In consistent with this re-population time, a careful analysis on PL decay of Figure 2B revealed that the residual PL signal at time delay > 100 ns decays with lifetime of 2 μs. This observation suggests additional complexity of the defect electronic states.

In summary, we show that strain, defect, layer engineering, as well as photon excitation energy, can be utilized together to manipulate the IX formation in $MoS_2/WSe_2$ heterostructures for robust quantum light generation in the 1.15 – 1.45 eV NIR spectral range. Our approach bypasses the momentum conservation limitation to form IX which otherwise would require lattice-matched component layers and perfect rotational alignment. In general, such an approach unlocks the freedom of combining arbitrary 2D semiconductors for building IXs with desired emission energy and extending this strategy for engineering interfacial $\Gamma$ point transitions in other 2D heterostructures, for instance, in $MoS_2/WS_2$. In consideration of the fact that monolayer TMDCs emit in the visible spectral range can be combined together to generate IXs that emit in the NIR, our findings pave the way for extending the operational wavelength of 2D quantum light sources into the technologically important yet rarely reached NIR regime.

## Materials and Methods

**Sample Preparation.** A detailed description of the sample preparation process can be found in Supplementary Section 1. All 2D flakes were exfoliated from bulk crystals purchased from HQ Graphene Inc.

**Optical Characterization.** A diagram of our optical measurement setup is presented in Supplementary Section 12. Micro-PL measurements were performed on a home-built confocal microscope with excitation from a supercontinuum pulsed laser with 77.6 MHz repetition rate unless otherwise specified. The excitation power was typically ~1 - 100 nW. Samples were mounted in a continuous flow cryostat and cooled to 10 K using liquid helium. The emitted light was collected through a 50× infrared objective lens (Olympus, 0.65 NA) and spectrally filtered before entering a 2D InGaAs array detector (NIRvana® LN, Princeton Instruments). We used 150 and 300 gr/mm gratings to resolve the spectra. For TRPL and HBT experiments, the emission signal was spectrally filtered before coupling into a 50:50 optical fiber beamsplitter, which equivalently split the signal into two beams and sent them into two channels of a superconducting nanowire single-photon detector (SNSPD, Quantum Opus). PL intensity time traces, PL decay curves, and $g^2(\tau)$ traces were obtained from photon detection events recorded by a PicoQuant HydraHarp 400 time-correlated single photon-counting module. For pulsed auto-correlation measurements, $g^2(0)$ was extracted by comparing the integrated photon coincidence counts at the zero-time delay peak with the averaged integrated photon coincidence counts at adjacent peaks.

**First-principles Calculations.** First principles calculations based on density functional theory (DFT) were performed within the general gradient approximate (GGA) using the Perdew-Burke-Ernzerhof (PBE) exchange-correlation functional as implemented in the Vienna Ab initio Simulation Package (VASP). The plane-wave energy cutoff is set to be 450 eV, and spin-orbit coupling is considered in our calculations. The vdW interactions in the heterostructures are included via the semiempirical Grimme-D3 scheme. A rigid shift of 0.92 eV is used to align with the experimental optical band gap as DFT is known to underestimate the band gap. For $MoS_2/WSe_2$ heterobilayer without defect, a k-grid of 15x15x1 is adopted. For defect systems, a single vacancy is included in a 4x4x1 supercell, where the k-grid sampling used is 3x3x1. The strain effect of the nanopillars is simulated by a biaxial tensile strain of 2%.

# Acknowledgements

**Funding:** This work was performed at the Center for Integrated Nanotechnologies, an Office of Science User Facility operated for the U.S. Department of Energy (DOE) Office of Science. Los Alamos National Laboratory (LANL), an affirmative action equal opportunity employer, is managed by Triad National Security, LLC for the U.S. Department of Energy's NNSA, under contract 89233218CNA000001. Deterministic quantum emitter creation capability was developed under the support of DOE BES, QIS Infrastructure Development Program BES LANL22. HH, XL, VC, and AP acknowledge partial support form Laboratory Directed Research and Development (LDRD) program 20200104DR. HZ and HH is also partially supported by Quantum Science Center. MTP is supported by LDRD 20210782ER and LDRD 20210640ECR. HZ also acknowledge a partial support from LANL Director's Postdoctoral Fellow Award. L.Z. is supported by the National Science Foundation (NSF) grant No. DMR-2124934, and L.Y. are supported by the Air Force Office of Scientific Research (AFOSR) grant No. FA9550-20-1-0255. This first-principles simulation uses the Extreme Science and Engineering Discovery Environment (XSEDE), which is supported by National Science Foundation (NSF) grant number ACI-1548562.
**Author Contributions:** HZ, under the supervision of HH, conceived, designed, and conducted the experiment, analyzed the data, and composed the initial draft of the paper. LZ and LY performed first principles calculations. MTP conducted Raman measurements. XL and VC assisted HZ in PL measurements. JKB assisted in sample preparation. AP performed the modeling of the TRPL and photon correlation function. HH and HZ wrote the manuscript with input from all the authors. **Competing Interests**: The authors declare no competing interests. **Data Availability**: All data needed to evaluate the conclusions in the paper are present in the paper and/or the Supplementary Materials. Additional data related to this paper may be requested from the authors.

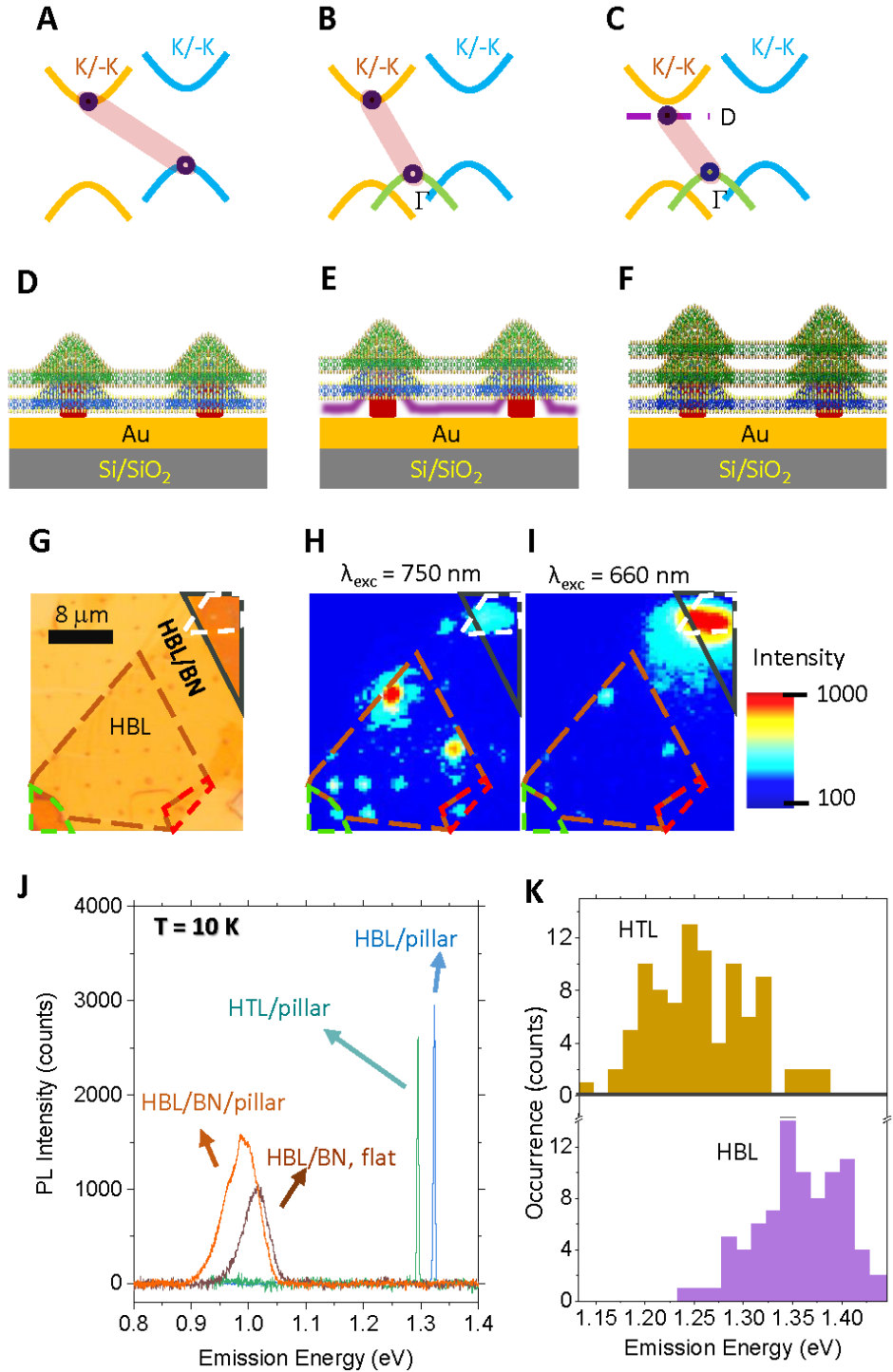

**Fig. 1. Interlayer excitons (IXs) from MoS$_2$-WSe$_2$ heterostructures.** (**A-C**) Type-II band alignment of K-K IX (**A**), Γ-K IX (**B**), and Γ-defect IX (**C**). The excitons are indicated by pink ovals. (**D-F**) Schematics of the heterostructures on nanopillar arrays (side view): HBL (**D**), HBL/hBN (**E**), and HTL (**F**). The green, blue, and purple layers are MoS$_2$, WSe$_2$, and hBN, respectively. The nanopillars residing on gold substrates are denoted by red color. (**G**) Optical image of an HBL flake on nanopillar arrays. Note that the top right corner is HBL/hBN. Brown

dashed line: HBL with 2° twist angle; white dashed line: HBL/hBN; black line: hBN flake; green line: 3L-MoS$_2$ / 1L-WSe$_2$; red line: 2L-MoS$_2$ / 1L-WSe$_2$ (HTL). (**H** and **I**) The wide-field PL image of **(G)**, under excitation wavelength of 750 nm and 660 nm, respectively. **(J)** Typical NIR PL spectra of HBL/nanopillar (blue), HTL/nanopillar, homogeneous HBL/hBN (brown), and HBL/hBN on nanopillar (red). **(K)** Histograms of emission energies of HBL (violet) and HTL (brown) on nanopillars. Data were collected in six different samples. All measurements were conducted at 10 K temperature.

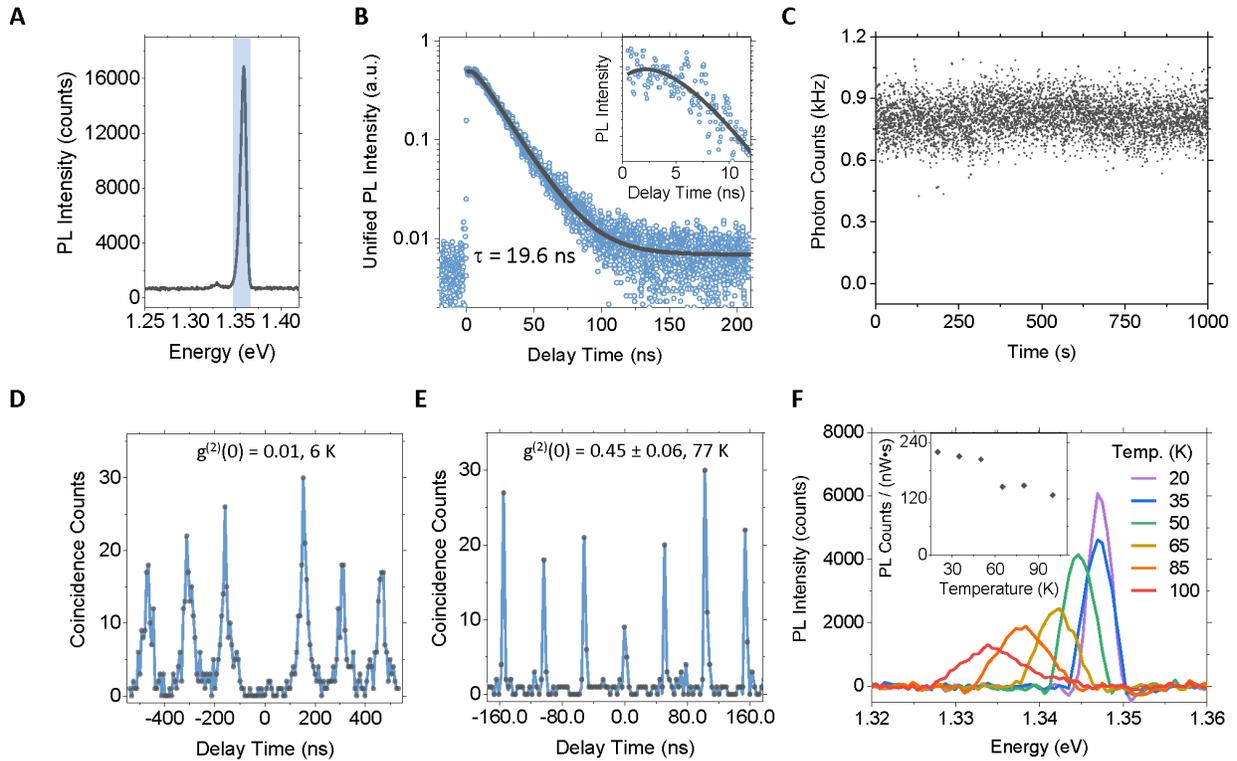

**Fig. 2. PL and photon correlation characterizations of localized HBL emitters on nanopillar arrays.** (**A**) PL spectrum of a localized HBL emitter on a nanopillar. The data in Fig. 2b-2d are taken from this emitter with a band-pass filter that allows the shadowed spectral region to be detected. (**B**) The PL decay curve (blue) and theoretical fit to the data (black) with an 18.45±0.01 ns extracted lifetime. Inset: The first 12 ns of the PL decay, which shows an uprising feature before the beginning of the single-exponential decay. (**C**) PL intensity as a function of experiment time showing stability of the single-photon emission rate over 1000 s time. (**D**) Second-order correlation measurement under 750 nm pulsed excitation with a 6.5 MHz repetition rate, from which a $g^{(2)}(0)$ = 0.01 is extracted. (**E**) Second-order correlation measurement of another HBL emitter on a nanopillar measured at 77 K, from which a $g^{(2)}(0)$ of 0.45±0.06 is extracted. (**F**) Temperature-dependent PL of a localized HBL emitter measured from 20 K to 100 K. The inset shows the evolution of integrated PL counts with temperature increase. All data were obtained at 10 K temperature unless otherwise specified.

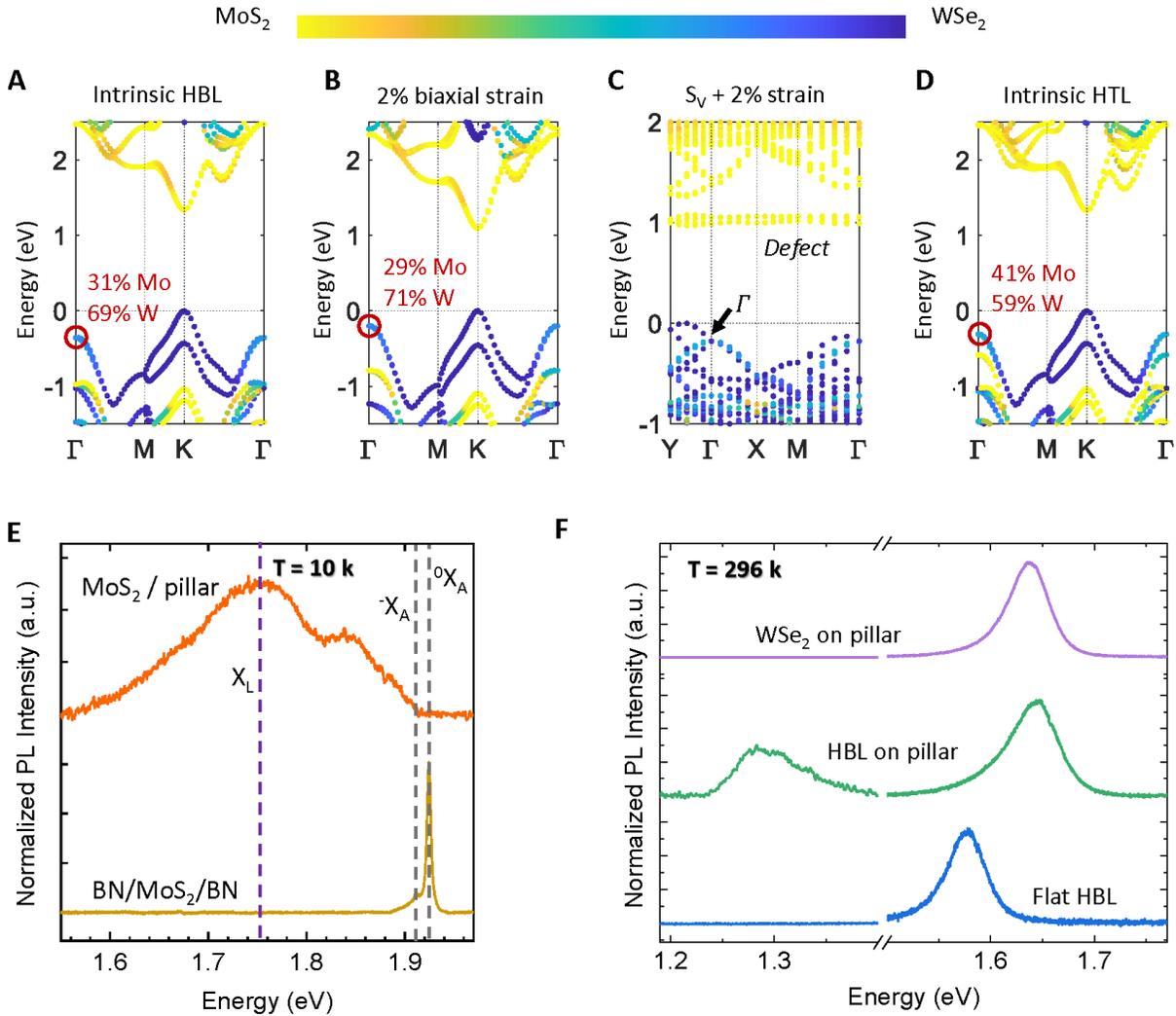

**Fig. 3. Understanding the roles of strain, defect, and layer-number engineering in MoS₂-WSe₂ IX emission.** (**A-D**) Band structures of (**A**) intrinsic MoS₂/WSe₂ HBL (**B**) MoS₂/WSe₂ HBL with 2% biaxial strain (**C**) MoS₂/WSe₂ HBL with a S vacancy and 2% biaxial strain (**D**) intrinsic 2L-MoS₂/1L-WSe₂ HTL. Projected wavefunction components of individual layers are represented by the color bar. The percentages of MoS₂ and WSe₂ wavefunction at Γ band edge are presented. (**E**) PL spectra of a homogeneous BN/MoS₂/BN sandwich (brown line) and a bare MoS₂ on nanopillar (orange line) measured at 10 K temperature. The exciton ($^0X_A$), trion ($^-X_A$), and defect-band ($X_L$) emission bands of MoS₂ are labeled. (**F**) Room-temperature PL spectra of a localized HBL emitter on nanopillar (green), a localized WSe₂ emitter on nanopillar (purple), and a homogeneous HBL sample (blue). The homogeneous HBL displays a Γ-K IX emission at 1.57 eV, which evinces a near-zero twist angle.

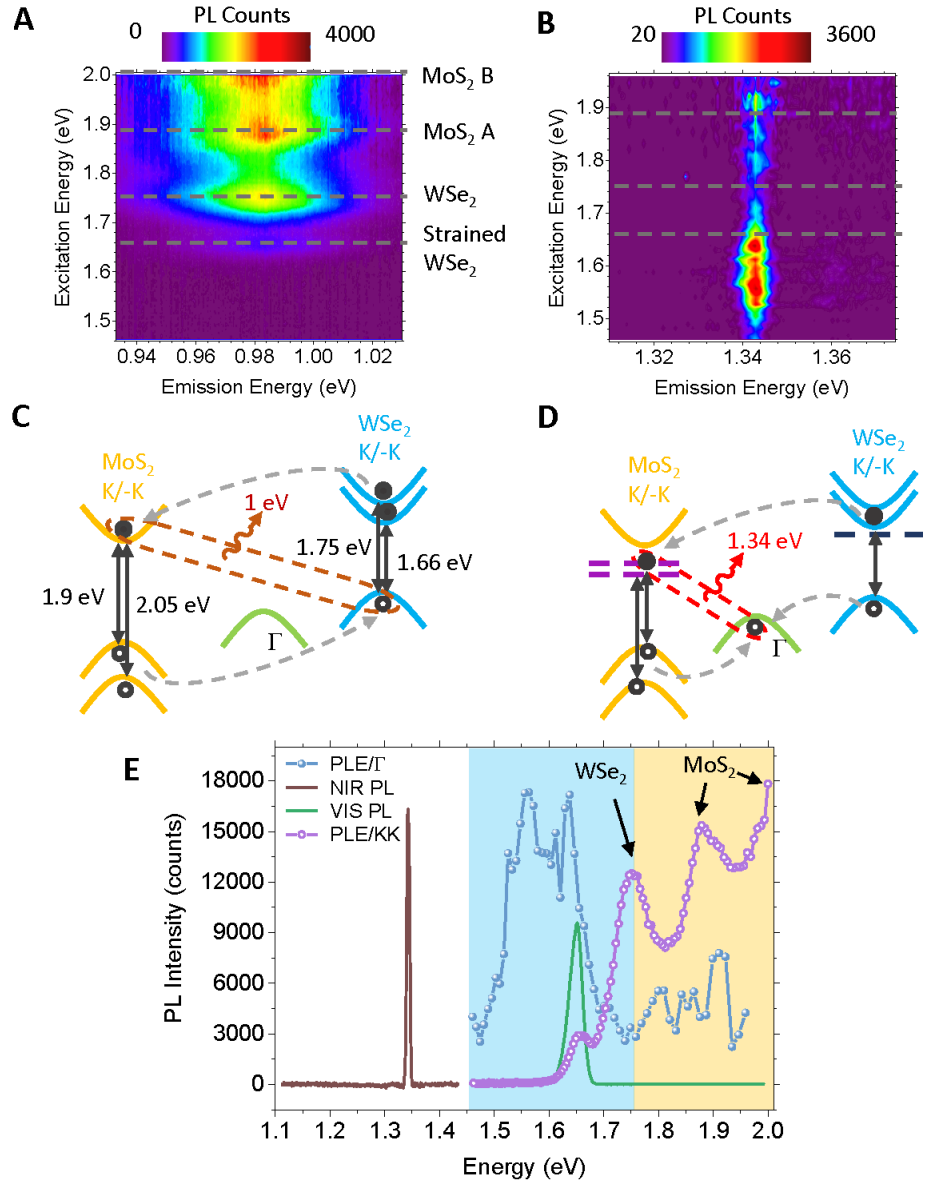

**Fig. 4. PLE study of MoS$_2$-WSe$_2$ IX emissions.** (**A** and **B**) The PLE intensity maps as a function of both excitation and emission photon energies. (**A**) the PLE map of a K-K IX emission on a nanopillar; (**B**) the PLE map of a Γ-defect IX emission on a nanopillar. The energy of the intralayer excitons of each material is marked by grey dashed lines. (**C** and **D**) Schematic representations of the energy band alignment in HBL/hBN on nanopillars (**C**), and HBL on nanopillars with defect bands (**D**). The black arrows show interband absorptions. The dashed gray arrows show charge transfer processes leading to the IX formations. The configuration for the K-K and Γ-defect IXs are represented by the dashed brown oval and the dashed red oval, respectively. (**E**) PL intensity at the K-K (purple line) and Γ-defect (blue line) IX peak as a function of excitation photon energy. The PL spectra of the same Γ-defect IX emitter collected at visible (green line) and near-infrared (brown line) spectral range are also shown. The spectral range of optical absorptions from WSe$_2$-branch and MoS$_2$-branch bound excitons are indicated by the blue and yellow shaded regions, respectively. All data were taken at 10 K temperature.